\documentclass[12pt,onecolumn]{IEEEtran}
\usepackage{dsfont}
\usepackage{setspace}
\usepackage{clrscode}
\usepackage{pict2e}

\usepackage{amsmath}
\usepackage{amssymb}
\usepackage[dvips]{graphicx}
\usepackage{epsfig}
\usepackage{algorithm}
\usepackage{algorithmic}
\usepackage{caption}
\usepackage{amsthm}
\usepackage{subcaption}

\usepackage[ps2pdf,
bookmarks=false,
bookmarksnumbered=false, 
bookmarksopen=false, 
colorlinks=false]{}

\makeatletter

\newcommand{\Rmnum}[1]{\expandafter\@slowromancap\romannumeral #1@}
\makeatother

\newcommand{\BS}{\text{BS}}
\newcommand{\UE}{\text{UE}}
\newcommand{\SINR}{\text{SINR}}

\newcommand{\g}{\text{g}}

\newcommand{\BE}{\text{BE}}

\newcommand{\Pout}{\text{P}_{\text{out}}}
\newcommand{\PoutL}{\text{P}_{\text{out},L}}
\newcommand{\PoutN}{\text{P}_{\text{out},N}}

\newcommand{\figsize}{.65}

\begin{document}
%
\title{Uplink Performance Analysis in D2D-Enabled mmWave Cellular Networks with Clustered Users}

\author{Esma Turgut and M. Cenk Gursoy
\thanks{The authors are with the Department of Electrical
Engineering and Computer Science, Syracuse University, Syracuse, NY, 13244
(e-mail: eturgut@syr.edu, mcgursoy@syr.edu).}}

\maketitle
\begin{spacing}{1.7}
\begin{abstract}
In this paper, an analytical framework is provided to analyze the uplink performance of device-to-device (D2D)-enabled millimeter wave (mmWave) cellular networks with clustered D2D user equipments (UEs). Locations of cellular UEs are modeled as a Poison Point Process (PPP), while locations of potential D2D UEs are modeled as a Poisson Cluster Process (PCP). Signal-to-interference-plus-noise ratio (SINR) outage probabilities are derived for both cellular and D2D links using tools from stochastic geometry. The distinguishing features of mmWave communications such as directional beamforming and having different path loss laws for line-of-sight (LOS) and non-line-of-sight (NLOS) links are incorporated into the outage analysis by employing a flexible mode selection scheme. Also, the effect of beamforming alignment errors on the outage probability is investigated to get insight on the performance in practical scenarios. Moreover, area spectral efficiency (ASE) of the entire network is determined for both underlay and overlay types of sharing. Optimal spectrum partition factor is determined for overlay sharing by considering the optimal weighted proportional fair spectrum partition.
\end{abstract}


\thispagestyle{empty}


\section{Introduction}
Recent years have witnessed an overwhelming increase in mobile data traffic due to e.g., ever increasing use of smart phones, portable devices, and data-hungry multimedia applications. Since limited available spectrum in microwave ($\mu$Wave) bands does not seem to be capable of meeting this demand in the near future, there has recently been significant interest in moving to new frequency bands. Therefore, the use of large bandwidth at millimeter wave (mmWave) frequency bands to provide much higher data rates and immense capacity has been proposed to be an important part of the fifth generation (5G) cellular networks and has attracted considerable attention recently
\cite{Rappaport1} -- \cite{Ghosh}.

Despite the great potential of mmWave bands, they have been considered attractive only for short range-indoor communication due to the increase in free-space path loss with increasing frequency, and poor penetration through solid materials. However, recent channel measurements and recent advances in RF integrated circuit design have motivated the use of these high frequencies for outdoor communication over a transmission range of about 150-200 meters \cite{Rappaport1}, \cite{Ghosh}. Also, with the employment of highly directional antennas, high propagation loss in the side lobes can be taken advantage of to support simultaneous communication with very limited or almost no interference to achieve lower link outage probabilities, much higher data rates and network capacity than those in $\mu$Wave networks.

Another promising solution to improve the network capacity is to enable device-to-device (D2D) communication in cellular networks. D2D communication allows proximity user equipments (UEs) to establish a direct communication link with each other
by bypassing the base station (BS). In other words, conventional two-hop cellular link is replaced by a direct D2D
link to enhance the network capacity. Network performance of D2D communication in cellular networks has recently been extensively studied as an important component of fourth generation (4G) cellular networks by using stochastic geometry. In \cite{Hesham1} and \cite{Ghosh2}, outage and spectrum efficiency of D2D-enabled uplink cellular networks were studied by considering mode selection schemes along with truncated channel inversion power control. In \cite{Ghosh2}, a distance-based mode selection scheme was employed while \cite{Hesham1} considered a flexible mode selection scheme. Also, effect of spectrum sharing type on the performance was investigated in \cite{Ghosh2}.
In these works, locations of the transmitting potential D2D UEs were modeled using Poisson Point Processes (PPPs) while the receiving D2D UEs were assumed to be distributed within a circle around the transmitting D2D UE. However, in D2D networks, UEs are very likely to form clusters rather than being distributed uniformly in the network. Therefore, a more realistic spatial model has been considered in several recent studies by modeling the locations of the D2D UEs as Poisson Cluster Process (PCP) distributed \cite{Afshang}, \cite{Afshang2}, \cite{Afshang3}. In \cite{Afshang}, authors obtained expressions for the coverage probability and area spectral efficiency of an out-of-band D2D network. Performance of cluster-centric content placement in a cache-enabled D2D network was studied in \cite{Afshang2}, where the authors have considered a cluster-centric approach which optimizes the performance of the entire cluster rather than the individual D2D UEs. In-band D2D communication where the cellular and D2D networks coexist in the same frequency band was considered in \cite{Afshang3} by combining PCP with a Poisson Hole Process (PHP). In particular, D2D UE locations are modeled by a Hole Cluster Process (HCP). However, neither of these works on D2D communication has addressed transmission in mmWave frequency bands. Network performance of D2D communication in cellular networks has been gaining even more importance in 5G networks and it is expected to be an essential part of mmWave 5G cellular networks.

Several recent studies have also addressed the mmWave D2D communication. In \cite{Qiao}, authors considered two types of D2D communication schemes in mmWave 5G cellular networks: local D2D and global D2D communications. Local D2D communication is performed by offloading the traffic from the BSs, while global D2D communication is established with multihop wireless transmissions via BSs between two wireless devices associated with different cells. The authors in \cite{Qiao} also proposed a resource sharing scheme to share network resources among local D2D and global D2D communications by considering the unique features of mmWave transmissions. In \cite{Guizani}, authors have proposed a resource allocation scheme in mmWave frequency bands, which enables underlay D2D communications to improve the system throughput and the spectral efficiency. mmWave D2D multi-hop routing for multimedia applications was studied in \cite{Eshraghi} to maximize the sum video quality by taking into account the unique characteristics of the mmWave propagation. In \cite{Turgut}, we have studied the uplink performance of D2D-enabled mmWave cellular networks where the locations of both cellular and potential D2D UEs are modeled as a PPP. In other words, correlation among the locations of potential D2D UEs was not taken into account (and also beamsteering errors and area spectral efficient were not addressed in \cite{Turgut}).

In this work, we consider a single-tier uplink network in which the BSs and cellular UEs coexist with the potential D2D UEs. We model the locations of BSs and cellular UEs as independent homogeneous PPPs. Unlike previous works on mmWave D2D communication systems where the D2D UEs are assumed to be uniformly distributed in the network, we model the locations of potential D2D UEs as a PCP to provide a more appropriate and realistic model. Moreover, potential D2D UEs in the clusters can choose to operate in cellular and D2D mode according to a mode selection scheme. Although there is a higher possibility of operating in D2D mode due to closer distances between the UEs in the clusters, this mode selection strategy provides flexibility and generality in our analysis. Additionally, different from the previous studies on D2D communications, most of which consider only underlay or overlay types of sharing, we take into account both types of sharing strategies to show their impact on the performance of the mmWave D2D networks.

More specifically, our main contributions can be summarized as follows:
\begin{itemize}
\item We provide an analytical framework to analyze the uplink performance of D2D-enabled mmWave cellular networks with clustered UEs by using tools from stochastic geometry. In particular, cellular and potential D2D UEs can coexist in the same band, and the cellular UEs are distributed uniformly and potential D2D UEs form clusters in the network.

\item An expression for the probability of selecting the D2D mode for a potential D2D UE located in a cluster is derived by considering a flexible mode selection scheme. Laplace transform expressions for  both cellular and D2D interference links are obtained. Using these characterizations, we derive SINR outage probability expressions for both cellular and D2D links employing the modified LOS ball model for blockage modeling.

\item We investigate the effect of spectrum sharing type on SINR outage probability. The effect of LOS ball model parameters is also identified. Additionally, the impact of alignment errors on the SINR outage probability is investigated to get insight on the performance in practical scenarios.

\item Area spectral efficiency (ASE) of the entire network is determined for both underlay and overlay types of sharing. We have shown that an optimal value for the average number of simultaneously active D2D links, maximizing the ASE, exists and this optimal value is independent of cluster center density. Moreover, optimal spectrum partition factor is found for overlay sharing by considering the optimal weighted proportional fair spectrum partition.

\end{itemize}

The rest of the paper is organized as follows. In Section \ref{sec:system_model}, system model is introduced. Characterizations for the transmission strategies and interference models are provided in Section \ref{Transmission and Interference Characterizations}. In Section \ref{sec:Analysis of Uplink SINR Outage Probability}, uplink SINR outage probabilities for both cellular and D2D links are derived initially considering perfect beam alignment, and then in the presence of beamsteering errors. ASE is defined and analyzed in Section \ref{sec:Analysis of Area Spectral Efficiency} for the underlay and overlay types of sharing. In Section \ref{sec:Simulation and Numerical Results}, simulations and numerical results are presented to identify the impact of several system parameters on the performance metrics. Finally, conclusions and suggestions for future work are provided in Section \ref{sec:Conclusion}. Proofs are relegated to the Appendix.

\section{System Model} \label{sec:system_model}
In this section, the system model for D2D-communication-enabled mmWave cellular networks with clustered UEs is presented. We consider a single-tier uplink network, where BSs are spatially distributed according to an independent homogeneous PPP $\Phi_{B}$ with density $\lambda_{B}$ on the Euclidean plane. UEs are categorized as cellular UEs and potential D2D UEs. Cellular UEs are distributed according to an independent homogeneous PPP $\Phi_{CU}$ with density $\lambda_{CU}$, while potential D2D UEs are clustered around the cluster centers in which the cluster centers are also distributed according to an independent homogeneous PPP $\Phi_{C}$ with density $\lambda_{C}$. For instance, cellular UEs can be regarded as pedestrians or UEs in transit which are more likely to be uniformly distributed in the network, and therefore homogeneous PPP is a better choice for the modeling of such UEs. On the other hand, potential D2D UEs are located in high UE density areas, i.e. hotspots, and are expected to be closer to each other forming clusters, and thus PCP is a more appropriate and accurate model than a homogeneous PPP.

Cluster members, i.e. potential D2D UEs, are  assumed to be symmetrically independently and identically distributed (i.i.d.) around the cluster center. The union of cluster members' locations form a PCP, denoted by $\Phi_{D}$. In this paper, we model $\Phi_D$ as a Thomas cluster process, where the UEs are scattered around the cluster center $x \in \Phi_C$ according to a Gaussian distribution with variance $\sigma_d^2$ and the probability density function (pdf) of a potential D2D UE's location is given by \cite{Haenggi}
\begin{equation}
f_{Y}(y)=\frac{1}{2\pi\sigma_d^2} \exp\left( -\frac{\|y\|^2 }{2\sigma_d^2}\right), \quad y \in \mathbb{R}^2.
\end{equation}
where $y$ is the UE's location relative to the cluster center and $\|\cdot\|$ is the Euclidean norm. Each potential D2D UE (i.e., each cluster member) in a cluster $x \in \Phi_C$ has the capability of establishing a direct D2D link with the cluster members in the same cluster or they can communicate with a BS in $\Phi_B$. Hence, potential D2D UEs can operate in one of the two modes according to the mode selection scheme: cellular and D2D mode. When operating in D2D mode, a potential D2D UE in the cluster bypasses the BS and communicates directly with its intended receiver in the same cluster. Let $\mathcal{N}^x$ denote the set of all potential D2D UEs in a cluster $x \in \Phi_C$, and $N$ be the total number of potential D2D UEs per cluster, which is assumed to be the same in each cluster. $\mathcal{N}^x$ can be divided into two subsets: set of possible transmitting potential D2D UEs ($\mathcal{N}_t^x$), and set of possible receiving D2D UEs ($\mathcal{N}_r^x$). The set of all simultaneously transmitting potential D2D UEs is denoted by $\mathcal{A}^x \subset \mathcal{N}_t^x$ where $|\mathcal{A}^x|$ is modeled as a Poisson distributed random variable with mean $\bar{n}$. $\mathcal{A}^x$ can also be divided into two subsets: set of simultaneously transmitting potential D2D UEs in D2D mode ($\mathcal{A}_d^x$) and set of simultaneously transmitting potential D2D UEs in cellular mode ($\mathcal{A}_c^x$) which are modeled as Poisson distributed random variables with means $\bar{n}P_{D2D}$ and $\bar{n}(1-P_{D2D})$, respectively. $P_{D2D}$ above is the probability of potential D2D UE selecting the D2D mode, and this probability will be described and characterized in detail later in the paper.

Without loss of generality, a typical receiving node (BS) is assumed to be located at the origin according to Slivnyak's theorem for cellular UEs and potential D2D UEs transmitting in cellular mode, and these UEs are assumed to be associated with their closest BS. The link between the BSs and cellular UEs/potential D2D UEs transmitting in cellular mode is called the cellular link, and the link between the transmitting and receiving D2D UEs in the same cluster is called the D2D link in the rest of the paper. For the D2D link, we conduct an analysis for a typical D2D UE located at the origin, which is randomly chosen in a randomly chosen cluster which is referred to as the representative cluster centered at $x_0 \in \Phi_C$ throughout the paper.

\section{Transmission Strategies and Interference Characterizations}\label{Transmission and Interference Characterizations}
In this section, we provide characterizations for the transmission strategies and interference models. In particular, we describe two types of spectrum sharing policies between the cellular and D2D UEs, identify the interference experienced in cellular uplink and D2D links, and  characterize the distributions of the link distances. Furthermore, we discuss the mode selection strategy and specify the beamforming assumptions.

\subsection{Spectrum Sharing}
Cellular spectrum can be shared between the cellular and D2D UEs in two different ways: underlay and overlay. In the underlay type of sharing, D2D UEs can opportunistically access the channel occupied by the cellular UEs. While for the overlay type of sharing, the uplink spectrum is divided into two orthogonal portions, i.e., a fraction $\delta$ of the cellular spectrum is assigned to D2D mode and the remaining $(1-\delta)$ fraction is used for cellular communication, where $\delta$ is the spectrum partition factor \cite{Ghosh2}. Also, parameter $\beta$ is defined as the spectrum sharing indicator which is equal to one for underlay and zero for overlay type of sharing.

\subsection{Interference Modeling}
\subsubsection{Interference in cellular uplink} The total interference in a cellular uplink experienced by a typical receiving node, i.e. the BS located at the origin, emerges from two sources: 1) interference from other cellular UEs/potential D2D UEs transmitting in cellular mode and 2) interference from other potential UEs transmitting in D2D mode (if underlay type of spectrum sharing is employed). Each cellular UE/potential D2D UE transmitting in cellular mode is assigned a unique and orthogonal channel by its associated BS which means that there is no intra-cell interference between UEs transmitting in cellular mode in the same cell. However, we assume universal frequency reuse across the entire cellular network causing inter-cell interference from the other cells' cellular UEs. Moreover, we consider a congested network scenario in which the total density of cellular UEs and potential D2D UEs in cellular mode is much higher than the density of BSs. In other words, each BS will always have at least one cellular UE to serve in the uplink channel. Different from the downlink communication, where we can model the interfering cellular UEs and potential D2D UEs in cellular mode in different cells as a PPP with density $\lambda_B$, modeling of the cellular interference for uplink is much more complicated \cite{Dhillon}. For example, an interfering UE in uplink case can be arbitrarily close to the BS, i.e., it can be closer than the UE being served. Therefore, one commonly used approach is to model the uplink other cell interferers as a non-homogeneous PPP $\Phi_c$ with a radially symmetric distance dependent density function given by
\begin{equation}
\lambda_u(t)=\lambda_{u,L}(t)+\lambda_{u,N}(t)=\sum_{j \in \{L,N\}} \lambda_B p_{j,c}(t)Q(t^{\alpha_{j,c}}) \label{density_function}
\end{equation}
where $Q(y)$ is the probability that path loss of a cellular UE to its serving BS is smaller than $y^{-1}$ \cite{Andrews4}. In the underlay case, we focus on one uplink channel which is shared by the cellular and D2D UEs. Since the potential D2D UEs operating in D2D mode coexist with the cellular UEs in an uplink channel, they cause both intra-cell and inter-cell interference at the BSs. On the other hand, in the overlay case, since the uplink spectrum is divided into two orthogonal portions, there is no cross-mode interference, i.e., no interference from the D2D UEs to the cellular UEs and vice versa.

\subsubsection{Interference in D2D link}
The total interference experienced by a typical D2D UE $\in \mathcal{N}_r^{x_0}$ in the representative cluster originates from three different sources: 1) cross-mode interference caused by the other cellular UEs/potential D2D UEs transmitting in cellular mode (if underlay sharing is adopted); 2) intra-cluster interference caused by the simultaneously transmitting D2D UEs in D2D mode inside the representative cluster; and 3) inter-cluster interference caused by the simultaneously transmitting D2D UEs in D2D mode outside the representative cluster. In the overlay case, there is no cross-mode interference, i.e., no interference from the cellular UEs/potential D2D UEs transmitting in cellular mode to the D2D UEs.

\subsection{Path-loss exponents and link distance modeling} \label{sec:Path-loss exponents and link distance modeling}
A transmitting UE can either have a line-of-sight (LOS) or non-line-of-sight (NLOS) link to the BS or the receiving UE. In a LOS state, UE should be visible to the receiving nodes, indicating that there is no blockage in the link. On the other hand, in a NLOS state, blockage occurs in the link. Consider an arbitrary link of length $r$, and define the LOS probability function $p(r)$ as the probability that the link is LOS. Using field measurements and stochastic blockage models, $p(r)$ can be modeled as $e^{-\zeta r}$ where decay rate $\zeta$ depends on the building parameter and density \cite{Bai1}. For analytical tractability, LOS probability function $p(r)$ can be approximated by a step function. In this approach, the irregular geometry of the LOS region is replaced with its equivalent LOS ball model. In this paper, modified LOS ball model is adopted similarly as in \cite{Andrews3}. According to this model, the LOS probability function of a link $p_L(r)$ is equal to some constant  $p_{L}$ if the link distance $r$ is less than ball radius $R_B$ and zero otherwise. The parameters $p_{L}$ and $R_B$ depend on geographical regions. $(p_{L,c}, R_{B,c})$ and $(p_{L,d}, R_{B,d})$ are the LOS ball model parameters for cellular and D2D links, respectively\footnote{Throughout the paper, subscripts $c$ and $d$ denote associations with cellular and D2D links, respectively.}. Therefore, LOS and NLOS probability function for each link can be expressed as follows:
\begin{align}\label{LOS_prob_funct}
  p_{L,\kappa}(r) &= p_{L,\kappa}\mathds{1}(r \le R_{B,\kappa}) \nonumber \\
  p_{N,\kappa}(r) &= (1-p_{L,\kappa})\mathds{1}(r \le R_{B,\kappa})+\mathds{1}(r > R_{B,\kappa})
\end{align}
for $\kappa \in \{c, d\}$ where $\mathds{1}(\cdot)$ is the indicator function. Different path loss laws are applied to LOS and NLOS links, and thus $\alpha_{L,\kappa}$ and $\alpha_{N,\kappa}$ are the LOS and NLOS path-loss exponents for $\kappa \in \{c, d\}$, respectively.

\subsubsection{D2D communication}
\begin{figure}
\centering
  \includegraphics[width=0.32\textwidth,angle=-90]{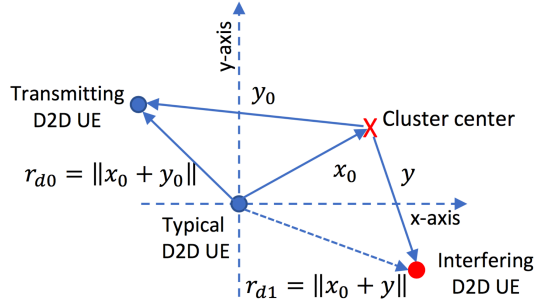}
  \caption{\small Illustration of the distances $r_{d0}$ and $r_{d1}$ in the representative cluster. The typical D2D UE is assumed to be located at the origin. Cluster center is located at $x_0$ with respect to (w.r.t.) the origin. Transmitting D2D UE is located at $y_0$ w.r.t. the cluster center. Intra-cluster interfering D2D UEs are located at $\{y\}$ w.r.t. the cluster center (Only one of them is shown in the figure). Arrows represent the coordinate vectors (and do not indicate the direction of communication).  \normalsize}
\label{Fig_Rep_Clu}
\end{figure}

Regarding the distance modeling for potential D2D UEs which are assumed to be located inside the clusters, there are three types of distances: 1) D2D link distance, i.e., serving distance, 2) intra-cluster interferer distances and 3) inter-cluster interferer distances. Without loss of generality, a typical receiving D2D UE $\in \mathcal{N}_r^{x_0}$ is assumed to be located at the origin, and is associated with another D2D UE $\in \mathcal{A}_d^{x_0}$ located at $y_0$ chosen uniformly at random within the same cluster. Note that the cluster center location is $x_0$ with respect to the origin (where the typical receiving D2D UE is), and transmitting D2D UE location is $y_0$ with respect to the cluster center. Fig. \ref{Fig_Rep_Clu} illustrates the considered setting where the relative locations are denoted by vectors. Also, let $r_{d0}=\|x_0+y_0\|$ denote the distance between the transmitting and typical receiving D2D UEs. Similarly, let $\{r_{d1}=\|x_0+y\|, \forall y \in \mathcal{A}_d^{x_0} \setminus y_0\}$ denote the set of the distances from simultaneously transmitting D2D UEs in D2D mode inside the representative cluster to a typical receiving D2D UE $\in \mathcal{N}_r^{x_0}$. Distances $r_{d0}$ and $r_{d1}$ are also illustrated in Fig. \ref{Fig_Rep_Clu}. Note that, $r_{d0}$ is the serving distance, and $\{r_{d1}\}$ is the set of distances from intra-cluster interfering D2D UEs. Actually, these distances are correlated due to the common factor $x_0$. By conditioning on $\omega_0=\|x_0\|$ and using the fact that $y_0$ and $\{y\}$ are i.i.d. zero-mean Gaussian random variables with variance $\sigma_d^2$ in $\mathbb{R}^2$, the serving distance $r_{d0}=\|x_0+y_0\|$ and the the set of distances from intra-cluster interfering D2D UEs $\{r_{d1}=\|x_0+y\|, \forall y \in \mathcal{A}_d^{x_0} \setminus y_0\}$ are conditionally i.i.d. It is shown that conditioning on $\omega_0$ instead of $x_0$ is sufficient \cite{Afshang}. Therefore, the pdf of each distance is characterized by a Rician distribution \cite{Afshang}:
\begin{align}
f_{R_{d0}}(r_{d0}|\omega_0)= \text{Ricepdf}(r_{d0},\omega_0;\sigma_d^2) \label{pdfofrd0} \\
f_{R_{d1}}(r_{d1}|\omega_0)= \text{Ricepdf}(r_{d1},\omega_0;\sigma_d^2) \label{pdfofrd1}
\end{align}
where $\text{Ricepdf}(a,b;\sigma_d^2)=\frac{a}{\sigma_d^2}\exp(-\frac{a^2+b^2}{2\sigma_d^2})I_0(\frac{ab}{\sigma_d^2})$ and $I_0(\cdot)$ is the modified Bessel function of the first kind with order zero. Similarly, let $\{r_{d2}=\|x+y\|, \forall y \in \mathcal{A}_d^{x}\}$ denote the set of the distances from simultaneously transmitting D2D UEs in D2D mode in the other clusters to a typical D2D UE $\in \mathcal{N}_r^{x_0}$, i.e., $\{r_{d2}\}$ is the set of distances from inter-cluster interfering D2D UEs. By conditioning on $\omega=\|x\|$, the pdf of each distance is given by $f_{R_{d2}}(r_{d2}|\omega)= \text{Ricepdf}(r_{d2},\omega;\sigma_d^2)$.

\subsubsection{Cellular communication} \label{sec:cellular link distance modeling}
Recall that cellular UEs and potential D2D UEs transmitting in cellular mode are assumed to be associated with their closest BS, and therefore the pdf of the cellular link distance $r_c$ to the nearest LOS/NLOS BS is given by \cite{Bai2}
\begin{equation}\label{cellularlinkpdftonearestBS}
f_s(r_c)= 2\pi\lambda_B r_c p_{s,c}(r_c) e^{-2\pi\lambda_B\psi_s(r_c)}/\mathcal{B}_{s,c} \quad \text{for} \; s \in \{L,N\}
\end{equation}
where $\psi_s(r_c)=\int_0^{r_c} x p_{s,c}(x) dx$, $\mathcal{B}_{s,c}=1-e^{-2\pi\lambda_B\int_0^{\infty} x p_{s,c}(x) dx}$ is the probability that a UE has at least one LOS/NLOS BS, and  $p_{s,c}(x)$ is given in (\ref{LOS_prob_funct}) for $s \in \{L,N\}$. In fact, since potential D2D UEs are distributed according to a PCP around the cluster centers, modeling the distance of potential D2D UEs transmitting in cellular mode to the closest BS by (\ref{cellularlinkpdftonearestBS}) is only an approximation. However, we have verified in the simulations that this assumption is quite reasonable especially for small values of the scattering variance $\sigma_d^2$.

Let $\{r_{y_x}=\|x+y\|, \forall x \in \Phi_C, \forall y \in \mathcal{A}_d^{x}\}$ be the set of distances from the cross-mode interferers, i.e. D2D UEs, to a typical BS at the origin. Then, the pdf of each distance is given by $f_{R_{y_x}}(r_{y_x}|\omega)= \text{Ricepdf}(r_{y_x},\omega;\sigma_d^2)$ where $\omega=\|x\|$.

\subsection{Mode Selection}
In this work, a flexible mode selection scheme similarly as in \cite{Hesham1} is considered. In this scheme, a potential D2D UE chooses the D2D mode if the biased D2D link quality is at least as good as the cellular uplink quality. In other words, a potential D2D UE will operate in D2D mode if $T_d r_d^{-\alpha_{s,d}} \geq  r_c^{-\alpha_{s,c}}$, where $T_d \in [0, \infty)$ is the biasing factor, and $r_c$ and $r_d$ are the cellular and D2D link distances, respectively. Biasing factor $T_d$ has two extremes, $T_d=0$ and $T_d \to \infty$. In the first extreme case, D2D communication is disabled, while in the second case, each potential D2D UE is forced to select the D2D mode. The probability of selecting D2D mode, $P_{D2D}$, is provided in the following Lemma.

\emph{Lemma 1:} Probability of selecting D2D mode for a potential D2D UE located in a cluster $x \in \Phi_C$ is
\begin{align}
P_{D2D}&= \sum_{s \in \{L,N\}}  \sum_{s^{\prime} \in \{L,N\}} \int_0^{\infty} \int_0^{\infty}  e^{-2\pi\lambda_B \psi_s\left(r_d^{\alpha_{s^{\prime},d}/\alpha_{s,c}}/T_d^{1/\alpha_{s,c}}\right)} f_{R_d}(r_d|\omega) f_{\Omega}(\omega) p_{s^{\prime},d}(r_d) dr_d d\omega
\end{align}
where $\psi_s(a)=\int_0^{a} x p_{s,c}(x) dx$, $p_{s,c}(x)$ and $p_{s^{\prime},d}(r_d)$ are given in (\ref{LOS_prob_funct}), $f_{R_d}(r_d|\omega)=\text{Ricepdf}(r_{d},\omega;\sigma_d^2)$, and $f_{\Omega}(\omega)=\frac{\omega}{\sigma_d^2}\exp(-\frac{\omega^2}{2\sigma_d^2})$.

\emph{Proof:} See Appendix \ref{Proof of Lemma 1}.

\subsection{Directional beamforming} \label{sec:Directional beamforming}
Antenna arrays at the BSs and UEs are assumed to perform directional beamforming where the main lobe being directed towards the dominant propagation path while smaller side lobes direct energy in other directions. For tractability in the analysis and similar to  \cite{Bai2}, \cite{Bai3}, \cite{Wildman}, \cite{Marco2}, antenna arrays are approximated by a sectored antenna model \cite{Hunter}. The array gains are assumed to be constant $M_{\nu}$ for all angles in the main lobe and another smaller constant $m_{\nu}$ in the side lobe for $\nu \in \{\BS, \UE\}$. Initially, perfect beam alignment \footnote{Subsequently, beamsteering errors are also addressed.} is assumed in between the transmitting nodes (e.g., cellular or potential D2D UEs) and receiving nodes (e.g., BSs or receiving D2D UEs), leading to an overall antenna gain of $M_{\BS}M_{\UE}$ for cellular link and $M_{\UE}M_{\UE}$ for D2D link. In other words, maximum directivity gain can be achieved for the intended link by assuming that the transmitting node and receiving node can adjust their antenna steering orientation using the estimated angles of arrivals. Also, the beam direction of the interfering nodes is modeled as a uniform random variable on $[0,2\pi)$. Therefore, the effective antenna gain is a discrete random variable (RV) described by

\begin{equation}
    G=\left\{
                \begin{array}{ll}
                  M_{l}M_{\UE} &\text{w. p.} \; p_{M_{l}M_{\UE}}=\frac{\theta_{l}}{2\pi} \frac{\theta_{\UE}}{2\pi} \\
                  M_{l}m_{\UE} &\text{w. p.} \; p_{M_{l}m_{\UE}}=\frac{\theta_{l}}{2\pi} \frac{2\pi-\theta_{\UE}}{2\pi} \\
                  m_{l}M_{\UE} &\text{w. p.} \; p_{m_{l}M_{\UE}}=\frac{2\pi-\theta_{l}}{2\pi} \frac{\theta_{\UE}}{2\pi}  \\
                  m_{l}m_{\UE} &\text{w. p.} \; p_{m_{l}m_{\UE}}=\frac{2\pi-\theta_{l}}{2\pi} \frac{2\pi-\theta_{\UE}}{2\pi}
                \end{array}
              \right. \label{eq:antennagains}
\end{equation}
for $l \in \{\BS, \UE\}$ where $\theta_{\nu}$ is the beam width of the main lobe for $\nu \in \{\BS, \UE\}$, and $p_{G}$ is the probability of having an antenna gain of $G$.

\vspace{-0.06cm}

\section{Analysis of Uplink SINR Outage Probability} \label{sec:Analysis of Uplink SINR Outage Probability}
In this section, we first develop a theoretical framework to analyze the uplink SINR outage probability for a generic UE operating in cellular mode or D2D mode using stochastic geometry. Although a biasing-based mode selection scheme is considered for selecting between D2D and cellular modes, the developed framework can also be applied to different mode selection schemes.
\subsection{Signal-to-Interference-plus-Noise Ratio (SINR)}
Recall that, without loss of generality, we consider a typical receiving node (BS or D2D UE $\in \mathcal{N}_r^{x_0}$ in the representative cluster) located at the origin. Therefore, the SINR experienced at a typical receiving node in cellular and D2D modes, respectively, can be written as
\begin{equation}
SINR^{c}=\frac{P_{c}G_0h_0r_c^{-\alpha_{c}(r_c)}}{\sigma_N^2+\underbrace{\sum_{i \in \Phi_c} P_cG_ih_ir_i^{-\alpha_{c}(r_i)}}_{I_{cc}}+\underbrace{\sum_{x \in \Phi_C} \sum_{y \in \mathcal{A}_d^{x}} P_d G_{y_x}h_{y_x}r_{y_x}^{-\alpha_{d}(r_{y_x})}}_{I_{dc}}}
\end{equation}
\begin{equation}
SINR^{d}=\frac{P_{d}G_0h_0r_{d0}^{-\alpha_{d}(r_{d0})}}{\sigma_N^2+\underbrace{\sum_{i \in \Phi_c} P_cG_ih_ir_i^{-\alpha_{c}(r_i)}}_{I_{cd}}+\underbrace{ \sum_{y \in \mathcal{A}_d^{x_0} \setminus y_0} P_d G_{y_{x_0}}h_{y_{x_0}}r_{d1}^{-\alpha_{d}(r_{d1})}}_{I_{dd_{\text{intra}}}}+\underbrace{\sum_{x \in \Phi_C \setminus x_0} \sum_{y \in \mathcal{A}_d^{x}} P_d G_{y_x}h_{y_x}r_{d2}^{-\alpha_{d}(r_{d2})}}_{I_{dd_{\text{inter}}}}}
\end{equation}
where $P_{\kappa}$ is the transmit power of the UE operating in mode $\kappa \in \{c,d\}$, $G_0$ is the effective antenna gain of the link which is assumed to be equal to $M_{\BS}M_{\UE}$ for cellular link and $M_{\UE}M_{\UE}$ for D2D link, $h_0$ is the small-scale fading gain, $\alpha_{\kappa}(\cdot)$ is the path-loss exponent of the link, which depends on whether the link is LOS on NLOS, $r_c$ and $r_{d0}$ are the cellular and D2D link distances, respectively, $\sigma_N^2$ is the variance of the additive white Gaussian noise component, $I_{c\kappa}$ is the aggregate interference at the receiving node from cellular UEs using the same uplink channel in different cells,  which constitute a non-homogeneous PPP $\Phi_c$ with density function given in (\ref{density_function}), and $I_{d\kappa}$ is the aggregate interference at the receiving node from D2D UEs located inside the clusters (hence including both inter-cell and intra-cell D2D UEs). For the D2D link, $I_{dd}$ has two components: intra-cluster interference $I_{dd_{\text{intra}}}$ and inter-cluster interference $I_{dd_{\text{inter}}}$. A similar notation is used for $I_{cc}$, $I_{cd}$, $I_{dc}$, $I_{dd_{\text{intra}}}$ and $I_{dd_{\text{inter}}}$, but note that the effective antenna gains $G_i$, $G_{y_x}$ and $G_{y_{x_0}}$, and path loss exponents $\alpha_{\kappa}(\cdot)$ are different for different interfering links as described in Section \ref{sec:Directional beamforming} and Section \ref{sec:Path-loss exponents and link distance modeling}, respectively.
Small-scale fading gains denoted by $h$ are assumed to have an independent exponential distribution in all links.

The above description implicitly assumes underlay spectrum sharing between cellular and D2D UEs. Note that since there is no cross-mode interference in the overlay case, the SINR expression in this case reduces to $SINR^{c}=\frac{P_{c}G_0h_0r_c^{-\alpha_{c}(r_c)}}{\sigma_N^2+I_{cc}}$, and $SINR^{d}=\frac{P_{d}G_0h_0r_{d0}^{-\alpha_{d}(r_{d0})}}{\sigma_N^2+I_{dd_{\text{intra}}}+I_{dd_{\text{inter}}}}$, for mode $\kappa \in \{c,d\}$.

\subsection{Laplace Transform of Interferences}
Before conducting the outage probability analysis, we first provide the Laplace transform expressions for each interference component. The thinning property of Poisson processes can be employed to split the interference component $I_{\chi}$ for $\chi \in \{cc,dc,cd,dd_{\text{intra}},dd_{\text{inter}}\}$ into 8 independent PPPs or PCPs as follows:
\begin{align}
I_{\chi} &= I_{\chi,L} + I_{\chi,N} \nonumber \\
&=\sum_{G \in \big\{ \substack {M_{l}M_{\UE},M_{l}m_{\UE},\\ m_{l}M_{\UE},m_{l}m_{\UE}}\big \}} \sum_{j \in \{L,N\}}I_{{\chi,j}}^{G},  \label{eq:6PPP}
\end{align}
for $l \in \{\BS, \UE\}$ where $I_{\chi,L}$ and $I_{\chi,N}$ are the aggregate LOS and NLOS interferences, and $I_{\chi,j}^{G}$ denotes the interference for $j \in \{L,N\}$ with random antenna gain $G$ defined in (\ref{eq:antennagains}). According to the thinning theorem, each independent nonhomogeneous PPP has a density of $\lambda_B p_{j,c}(t)Q(t^{\alpha_{j,c}}) p_{G}$ for $\chi=\{cc,cd\}$, and each independent PCP has a density of $\lambda_C p_G$ for $\chi=\{dc,dd_{\text{inter}}\}$ where $p_{G}$ is given in (\ref{eq:antennagains}) for each antenna gain $G$. Similarly, for intracell interference on D2D link, i.e. $I_{dd_{\text{intra}}}$, number of interfering D2D UEs are thinned by multiplying $p_{G}$ with $\bar{n} P_{D2D}-1$ where $\bar{n} P_{D2D}$ is the mean number of simultaneously transmitting potential D2D UEs in D2D mode.

Inserting (\ref{eq:6PPP}) into the Laplace transform expression and using the definition of the Laplace transform yield
\begin{align}
\mathcal{L}_{I_{\chi}}(v)&= \mathbb{E}_{I_{\chi}}[e^{-vI_{\chi}}]=\mathbb{E}_{I_{\chi}}[e^{-v(I_{\chi,L}+I_{\chi,N})}] \nonumber \\
&\stackrel{(a)}{=}\mathbb{E}_{I_{\chi,L}}\left[e^{-v \sum_{G} I_{\chi,L}^{G}}\right] \times \mathbb{E}_{I_{\chi,N}}\left[{e^{-v \sum_{G} I_{\chi,N}^{G}}}\right] \nonumber \\
&= \prod_G \prod_j \mathbb{E}_{I_{\chi,j}^G}\left[ e^{-vI_{\chi,j}^G}\right] \nonumber \\
&=\prod_G \prod_j \mathcal{L}_{I_{\chi,j}^G}(v), \label{eq:LT}
\end{align}
where $G \in \{M_{l}M_{\UE},M_{l}m_{\UE}, m_{l}M_{\UE},m_{l}m_{\UE}\}$ for $l \in \{\BS, \UE\}$, $j \in \{L,N\}$, and (a) follows from the fact that  $I_{\chi,L}$ and $I_{\chi,N}$ are interferences generated from two independent thinned PPPs or PCPs.

Laplace transform expressions for each interference component are provided in the following Lemmas.

\emph{Lemma 2:} Laplace transform of the aggregate interference at the BS from cellular UEs using the same uplink channel in different cells is given by
\begin{align}
\mathcal{L}_{I_{cc}}(v)=\exp\left(-2\pi\lambda_B\sum_{j \in \{L,N\}}  \sum_{i=1}^4 p_{G_i} \int_{0}^{\infty} \frac{v P_cG_i t^{-\alpha_{j,c}}}{1+v P_cG_i t^{-\alpha_{j,c}}} Q\left(t^{\alpha_{j,c}}\right) p_{j,c}(t)tdt\right) \label{LT_cc}
\end{align}
where $v=\frac{\Gamma r_c^{\alpha_{s,c}}}{P_{c}G_0}$, $p_{j,c}(t)$ is the LOS/NLOS probability function for cellular link given in (\ref{LOS_prob_funct}), and $Q(y)$ is defined as \cite{Andrews4}
\begin{equation}
Q(y)=1-\exp\left(-2\pi\lambda_B\left( \int_{0}^{y^{1/\alpha_{L,c}}} x p_{L,c}(x) dx+ \int_{0}^{y^{1/\alpha_{N,c}}} x p_{N,c}(x) dx \right) \right) \label{Q(y)}
\end{equation}.

\emph{Proof:} See Appendix \ref{Proof of Lemma 2}.

\emph{Lemma 3:} Laplace transform of the aggregate interference at the BS from both intracell and intercell D2D UEs is given by
\begin{align}
&\mathcal{L}_{I_{dc}}(v)= \nonumber \\
& \exp\left(-2\pi\lambda_C  \sum_{j \in \{L,N\}}  \sum_{i=1}^4 p_{G_i} \int_0^{\infty} \left(1-\exp\left(-\bar{n}P_{D2D} \int_{0}^{\infty}\frac{v P_dG_i u^{-\alpha_{j,d}}}{1+v P_dG_i u^{-\alpha_{j,d}}}f_{U}(u|t)p_{j,d}(u)du\right)\right)tdt\right)
\end{align}
which can be approximated by
\begin{align}
\mathcal{L}_{I_{dc}}(v) \approx \exp\left(-2\pi\lambda_C \bar{n}P_{D2D} \sum_{j \in \{L,N\}}  \sum_{i=1}^4 p_{G_i} \int_0^{\infty} \frac{v P_dG_i u^{-\alpha_{j,d}}}{1+v P_dG_i u^{-\alpha_{j,d}}}p_{j,d}(u)udu\right)
\end{align}
where $v=\frac{\Gamma r_c^{\alpha_{s,c}}}{P_{c}G_0}$, $p_{j,d}(u)$ is the LOS/NLOS probability function for D2D link given in (\ref{LOS_prob_funct}), and the approximation follows from the Taylor series expansion of exponential function, i.e. $1-\exp(-x)\approx x$ for small $x$, and the Rician distribution property that $\int_{0}^{\infty}f_{U}(u|t)tdt=u$.

\emph{Proof:} See Appendix \ref{Proof of Lemma 3}.

\emph{Lemma 4:} Laplace transform of the aggregate interference at the typical UE from cellular UEs using the same uplink channel in different cells is given by
\begin{align}
\mathcal{L}_{I_{cd}}(v)=\exp\left(-2\pi\lambda_B\sum_{j \in \{L,N\}}  \sum_{i=1}^4 p_{G_i} \int_{0}^{\infty} \frac{v P_cG_i t^{-\alpha_{j,c}}}{1+v P_cG_i t^{-\alpha_{j,c}}}p_{j,c}(t)tdt\right) \label{LT_cd}
\end{align}
where $v=\frac{\Gamma r_{d0}^{\alpha_{s,d}}}{P_{d}G_0}$, $p_{j,c}(t)$ is the LOS/NLOS probability function for cellular link given in (\ref{LOS_prob_funct}), and $Q(y)$ is given in (\ref{Q(y)}).

\emph{Proof:} Proof follows similar steps as in the proof of Lemma 2.

\emph{Lemma 5:} Laplace transform of the intra-cluster interference at the typical UE $\in \mathcal{N}_r^{x_0}$ in the representative cluster is given by
\begin{align}
\mathcal{L}_{I_{dd_{\text{intra}}}}(v|w_0) = \exp \left(-\left(\bar{n}P_{D2D}-1\right)  \sum_{j \in \{L,N\}}  \sum_{i=1}^4 p_{G_i} \int_0^{\infty} \frac{v G_i u^{-\alpha_{j,d}}}{1+v G_i u^{-\alpha_{j,d}}}f_{U}(u|w_0)p_{j,d}(u)du \right)
\end{align}
where $v=\frac{\Gamma r_{d0}^{\alpha_{s,d}}}{P_dG_0}$, and $p_{j,d}(u)$ is the LOS/NLOS probability function for D2D link given in (\ref{LOS_prob_funct}).

\emph{Proof:} See Appendix \ref{Proof of Lemma 5}.

\emph{Lemma 6:} Laplace transform of the inter-cluster interference at the typical UE $\in \mathcal{N}_r^{x_0}$ in the representative cluster is given by
\begin{align}
&\mathcal{L}_{I_{dd_{\text{inter}}}}(v) = \nonumber \\
& \exp\left(-2\pi\lambda_C  \sum_{j \in \{L,N\}}  \sum_{i=1}^4 p_{G_i} \int_0^{\infty} \left(1-\exp\left(-\bar{n}P_{D2D} \int_{0}^{\infty}\frac{vP_d G_i u^{-\alpha_{j,d}}}{1+vP_d G_i u^{-\alpha_{j,d}}}f_{U}(u|t)p_{j,d}(u)du\right)\right)tdt\right)
\end{align}
which can be approximated by
\begin{align}
\mathcal{L}_{I_{dd_{\text{inter}}}}(v) \approx \exp\left(-2\pi\lambda_C \bar{n}P_{D2D} \sum_{j \in \{L,N\}}  \sum_{i=1}^4 p_{G_i} \int_0^{\infty} \frac{v P_dG_i u^{-\alpha_{j,d}}}{1+vP_d G_i u^{-\alpha_{j,d}}}p_{j,d}(u)udu\right)
\end{align}
where $v=\frac{\Gamma r_{d0}^{\alpha_{s,d}}}{P_dG_0}$.

\emph{Proof:} Proof follows similar steps as in the proof of Lemma 3.

\subsection{Uplink SINR Outage Probability}\label{sec:Uplink SINR Outage}
The uplink SINR outage probability $\Pout$ is defined as the probability that the received SINR is less than a certain threshold $\Gamma>0$, i.e., $\Pout= \mathbb{P}(\SINR<\Gamma)$. The outage probability for a typical UE in cellular mode is given in the following theorem.

\textit{Theorem 1:} In a single-tier D2D-communication-enabled mmWave cellular network, the outage probability for a typical cellular UE can be expressed as
\begin{align}
&\Pout^c(\Gamma)=1-\sum_{s \in \{L,N\}} \int_{0}^{\infty} e^{-\frac{ \Gamma r_c^{\alpha_{s,c}}\sigma_N^2}{P_{c}G_0}} \mathcal{L}_{I_{cc}}\left(\frac{\Gamma r_c^{\alpha_{s,c}}}{P_{c}G_0}\right) \mathcal{L}_{I_{dc}}\left(\frac{\beta \Gamma r_c^{\alpha_{s,c}}}{P_{c}G_0}\right) f_{s}(r_c) \mathcal{B}_{s,c} dr_c \label{eq:Pout_c}
\end{align}
where the Laplace transforms $\mathcal{L}_{I_{cc}}(v)$ and $\mathcal{L}_{I_{dc}}(\beta v)$ are given in Lemma 2 and Lemma 3, respectively, and $f_{s}(r_c)$ is the pdf of the cellular link distance given in (\ref{cellularlinkpdftonearestBS}).

\textit{Proof:} See Appendix \ref{Proof of Theorem 1}.

\textit{Theorem 2:} In a single-tier D2D-communication-enabled mmWave cellular network, the outage probability for a typical D2D UE can be expressed as
\begin{align}
\Pout^d(\Gamma)=1-&\sum_{s \in \{L,N\}} \int_{0}^{\infty} e^{-\frac{\Gamma r_{d0}^{\alpha_{s,d}}\sigma_N^2}{P_{d}G_0}} \mathcal{L}_{I_{dd_{\text{intra}}}}\left(\frac{ \Gamma r_{d0}^{\alpha_{s,d}}}{P_{d}G_0}|w_0\right) \mathcal{L}_{I_{dd_{\text{inter}}}}\left(\frac{ \Gamma r_{d0}^{\alpha_{s,d}}}{P_{d}G_0}\right)  \mathcal{L}_{I_{cd}}\left(\frac{\beta \Gamma r_{d0}^{\alpha_{s,d}}}{P_{d}G_0}\right) \nonumber \\
&\times p_{s,d}(r_{d0}) f_{R_{d0}}(r_{d0}|w_0)f_{\Omega_0}(w_0) dr_{d0}dw_0 \label{eq:Pout_d}
\end{align}
where the Laplace transforms $\mathcal{L}_{I_{dd_{\text{intra}}}}(v|w_0)$, $\mathcal{L}_{I_{dd_{\text{inter}}}}(v)$ and $\mathcal{L}_{I_{cd}}(\beta v)$ are given in Lemma 4, Lemma 5 and Lemma 6, respectively, $p_{s,d}(r_{d0})$ is the LOS/NLOS probability function for D2D link given in (\ref{LOS_prob_funct}), $f_{R_{d0}}(r_{d0}|w_0)$ is the pdf of the D2D link distance given in (\ref{pdfofrd0}), and $f_{\Omega_0}(\omega_0)=\frac{\omega_0}{\sigma_d^2}\exp(-\frac{\omega_0^2}{2\sigma_d^2})$.

\textit{Proof:} Proof follows similar steps as in the proof of Theorem 1, and the details are omitted for the sake of brevity.

\subsection{Uplink SINR Outage Probability Analysis In the Presence of Beamsteering Errors}
In Section \ref{sec:Uplink SINR Outage} and the preceding analysis, antenna arrays at the transmitting nodes (cellular or potential D2D UEs) and receiving nodes (BSs or UEs) are assumed to be aligned perfectly and uplink SINR outage probabilities are calculated in the absence of beamsteering errors. However, in practice, it may not be easy to have perfect alignment. Therefore, in this section, we investigate the effect of beamforming alignment errors on the outage probability analysis. We employ an error model similar to that in \cite{Wildman}. Let $|\epsilon|$ be the random absolute beamsteering error of the transmitting node toward the receiving node with zero-mean and bounded absolute error $|\epsilon|_{\text{max}} \le \pi$. Due to the symmetry in the gain $G_0$, it is appropriate to consider the absolute beamsteering error. The PDF of the effective antenna gain $G_0$ with alignment error can be explicitly written as \cite{Marco2}
\begin{align}
&f_{G_0}(\g)=F_{|\epsilon|}\left(\frac{\theta_{l}}{2}\right)F_{|\epsilon|}\left(\frac{\theta_{\UE}}{2}\right)\delta(\g-M_{l}M_{\UE})+F_{|\epsilon|}\left(\frac{\theta_{l}}{2}\right)\left(1-F_{|\epsilon|}\left(\frac{\theta_{\UE}}{2}\right)\right)
\delta(\g-M_{l}m_{\UE}) \nonumber \\
&+\left(1-F_{|\epsilon|}\left(\frac{\theta_{l}}{2}\right)\right)F_{|\epsilon|}\left(\frac{\theta_{\UE}}{2}\right) \delta(\g-m_{l}M_{\UE}) +\left(1-F_{|\epsilon|}\left(\frac{\theta_{l}}{2}\right)\right)\left(1-F_{|\epsilon|}\left(\frac{\theta_{\UE}}{2}\right)\right) \delta(\g-m_{l}m_{\UE}),
\label{eq:PDFofG}
\end{align}
for $l \in \{\BS, \UE\}$ where $\delta(\cdot)$ is the Kronecker's delta function, $F_{|\epsilon|}(x)$ is the CDF of the misalignment error and (\ref{eq:PDFofG}) follows from the definition of CDF, i.e., $F_{|\epsilon|}(x)=\mathbb{P}\{|\epsilon|\le x\}$. Assume that the error $\epsilon$ is Gaussian distributed, and therefore the absolute error $|\epsilon|$ follows a half normal distribution with $F_{|\epsilon|}(x)=\text{erf}(x/(\sqrt{2}\sigma_{\BE}))$, where $\text{erf}(\cdot)$ denotes the error function and $\sigma_{\BE}$ is the standard deviation of the Gaussian error $\epsilon$.

It is clear that the uplink SINR outage probability expressions in Section \ref{sec:Uplink SINR Outage} depend on the effective antenna gain $G_0$ between the transmitting and the receiving nodes. Thus, uplink SINR outage probability $\Pout^{\kappa}(\Gamma)$ for a typical UE in mode $\kappa \in \{c,d\}$ can be calculated by averaging over the distribution of $G_0$, $f_{G_0}(\g)$, as follows:
\begin{align}
\Pout^{\kappa}(\Gamma) &= \int_0^{\infty}\Pout^{\kappa}(\Gamma;g)f_{G_0}(\g)d \g \nonumber \\
&=F_{|\epsilon|}(\theta_{l}/2)F_{|\epsilon|}(\theta_{\UE}/2) \Pout^{\kappa}(\Gamma;M_{l}M_{\UE})+F_{|\epsilon|}(\theta_{l}/2) \bar{F}_{|\epsilon|}(\theta_{\UE}/2) \Pout^{\kappa}(\Gamma;M_{l}m_{\UE}) \nonumber \\
&+ \bar{F}_{|\epsilon|}(\theta_{\l}/2)F_{|\epsilon|}(\theta_{\UE}/2) \Pout^{\kappa}(\Gamma;m_{l}M_{\UE})+\bar{F}_{|\epsilon|}(\theta_{\l}/2) \bar{F}_{|\epsilon|}(\theta_{\UE}/2) \Pout^{\kappa}(\Gamma;m_{l}m_{\UE}),
\end{align}
for $l \in \{\BS, \UE\}$ where we define $\bar{F}_{|\epsilon|}(\theta/2)=1-F_{|\epsilon|}(\theta/2)$.

\section{Analysis of Area Spectral Efficiency} \label{sec:Analysis of Area Spectral Efficiency}
In Section \ref{sec:Analysis of Uplink SINR Outage Probability}, we have analyzed the uplink outage probability and obtained outage probability expressions for a typical cellular and D2D link. In this section, we consider another performance metric, namely area spectral efficiency (ASE), to measure the network capacity. ASE is defined as the average number of bits transmitted per unit time per unit bandwidth per unit area. It can be mathematically defined as follows:
\begin{equation}
\text{ASE}=\left(\lambda_B(1-\Pout^c(\Gamma))+\bar{n}P_{D2D} \lambda_C (1-\Pout^d(\Gamma))\right)\log_2(1+\Gamma)   \label{eq:ASE_underlay}
\end{equation}
where $\Pout^c(\Gamma)$ and $\Pout^d(\Gamma)$ are given in (\ref{eq:Pout_c}) and (\ref{eq:Pout_d}), respectively, $\bar{n}P_{D2D} \lambda_C$ and $\lambda_B$ are the average number of simultaneously active D2D links and cellular links per unit area, respectively. Note that ASE defined in (\ref{eq:ASE_underlay}) is valid for a saturated network scenario, i.e., each BS has at least one cellular UE to serve in the uplink channel. If the network is not saturated, the presence of inactive BSs will lead to increased SINR for both cellular and D2D links (due to lower interference), and outage probability will decrease. However, ASE may be lower as a result of fewer number of active cellular links per unit area. The ASE expression in (\ref{eq:ASE_underlay}) is given for underlay type of spectrum sharing. For overlay type of sharing, the uplink spectrum is divided into two orthogonal portions. Therefore, ASE can be redefined as follows:
\begin{equation}
\text{ASE}=\left((1-\delta)\lambda_B(1-\Pout^c(\Gamma))+\delta \bar{n}P_{D2D} \lambda_C (1-\Pout^d(\Gamma))\right)\log_2(1+\Gamma)
\end{equation}
where $\delta$ is the spectrum partition factor. In the case of overlay spectrum sharing, the following optimization problem can be formulated in order to determine the optimal value of $\delta$ that maximizes the ASE:
\begin{equation}
\delta^{\ast}=\arg \underset{\delta \in [0,1]}{\max} \; \text{ASE}
\end{equation}
The solution of this optimization problem is given as follows: if $\lambda_B(1-\Pout^c(\Gamma)) > \bar{n}P_{D2D} \lambda_C (1-\Pout^d(\Gamma))$, $\delta^{\ast}=0$; otherwise, $\delta^{\ast}=1$. In other words, all bandwidth is assigned to the cellular or D2D link depending on which one is performing better. Therefore, this is a greedy approach that does not address any fairness considerations. In numerical results, we have shown that if cellular communication is disabled, i.e. $\delta=1$, ASE is maximized. To overcome this unfairness in the spectrum allocation between D2D and cellular communication, we consider the optimal weighted proportional fair spectrum partition which is formulated as follows:
\begin{equation}
\delta^{\ast}=\arg \underset{\delta \in [0,1]}{\max} \; w_c\log\big((1-\delta)\lambda_B(1-\Pout^c(\Gamma))\log_2(1+\Gamma)\big)+w_d\log\left(\delta \bar{n}P_{D2D} \lambda_C (1-\Pout^d(\Gamma))\log_2(1+\Gamma)\right) \label{eq:objection_func}
\end{equation}
where $w_c$ and $w_d$ are the introduced weights. If we take the derivative of the objective function in (\ref{eq:objection_func}) with respect to $\delta$ and make it equal to zero, the optimal spectrum partition factor is obtained as $\delta^{\ast}=\frac{w_d}{w_c+w_d}=w_d$ which is simply equal to the weight we assign to the potential D2D UEs. In other words, $w_d$ portion of the spectrum should be assigned to D2D communication to achieve proportional fairness.

\section{Simulation and Numerical Results} \label{sec:Simulation and Numerical Results}
In this section, theoretical expressions are evaluated numerically. We also provide simulation results to validate the accuracy of the proposed model for the D2D-enabled uplink mmWave cellular network with clustered UEs as well as to confirm the accuracy of the analytical characterizations. In the numerical evaluations and simulations, unless stated otherwise, the parameter values listed in Table \ref{Table} are used.

\begin{table}
\small
\caption{System Parameters}
\centering
  \begin{tabular}{| l | r|}
    \hline
    \textbf{Parameters} & \textbf{Values}  \\ \hline
    $\alpha_{L,c}$, $\alpha_{N,c}$; $\alpha_{L,d}$, $\alpha_{N,d}$ & 2, 4; 2, 4\\ \hline
    $M_{\nu}$, $m_{\nu}$, $\theta_{\nu}$ for $\nu \in \{\BS, \UE\}$ & 20dB, -10dB, $30^o$  \\ \hline
    $\lambda_B$, $\lambda_C$ & $10^{-5}$, $10^{-4}$ $(1/m^2)$ \\ \hline
    $N$, $\bar{n}$ & 40, 3 \\ \hline
    $(p_{L,c}, R_{B,c})$, $(p_{L,d}, R_{B,d})$ & (1, 100), (1, 50) \\ \hline
    $\beta$, $\delta$, $T_d$ & 1, 0.2, 1 \\ \hline
    $\Gamma$, $\sigma_N^2$, $\sigma_d^2$  & 0dB, -74dBm, 25 \\ \hline
    $P_c$, $P_d$ & 200mW, 200mW  \\ \hline
  \end{tabular} \label{Table}
\end{table}

\begin{figure}
\centering
  \includegraphics[width=\figsize\textwidth]{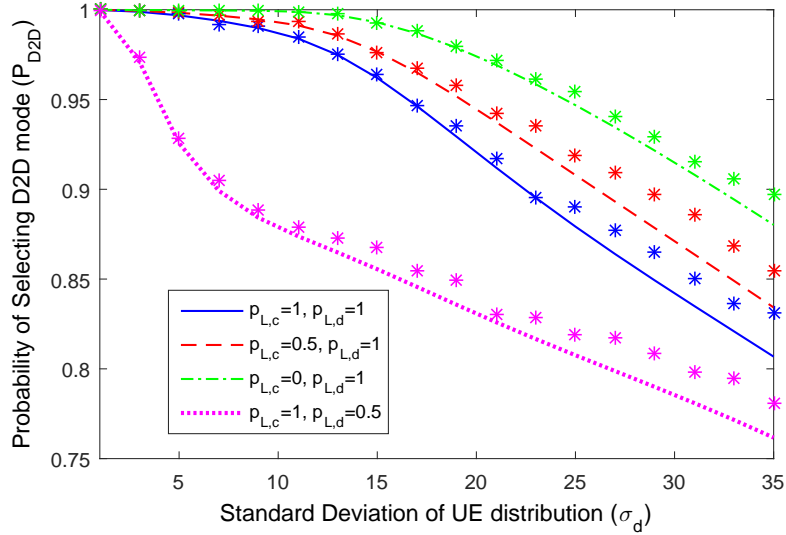}
  \caption{\small Probability of selecting D2D mode  as a function of UE distribution's standard deviation $\sigma_d$ for different values of $p_{L,c}$ and $p_{L,d}$. Simulation results are also plotted with markers. \normalsize}
\label{Fig_PD2D}
\end{figure}

First, we investigate the effect of UE distribution's standard deviation $\sigma_d$ on the probability of selecting D2D mode for different values of the LOS probability function $p_{L,c}$ for cellular link and $p_{L,d}$ for D2D link in Fig. \ref{Fig_PD2D}. As the standard deviation increases, the distance between the transmitting and receiving potential D2D UEs also increases. As a result, probability of selecting the D2D mode decreases. Also, since the number of LOS BSs increases with the increase in $p_{L,c}$, probability of selecting D2D mode decreases with increasing $p_{L,c}$. On the other hand, probability of selecting D2D mode increases when we increase $p_{L,d}$ as a result of increasing number of LOS potential D2D UEs in the cluster. As we have discussed in Section \ref{sec:cellular link distance modeling}, although the cluster centers are distributed according to a PPP, modeling the pdf of the distance between the nearest LOS/NLOS BS and potential D2D UE using Eq. (\ref{cellularlinkpdftonearestBS}) is only an approximation because the potential D2D UEs are distributed according to a PCP around the cluster centers. However, as shown in Fig. \ref{Fig_PD2D}, this pdf assumption agrees well with the simulation results especially for small values of $\sigma_d$. On the other hand, there is a minor deviation between the analysis and simulation results for larger values of $\sigma_d$. This is because potential D2D UEs are located farther from the cluster center for larger $\sigma_d$.

\begin{figure}
\centering
  \includegraphics[width=\figsize\textwidth]{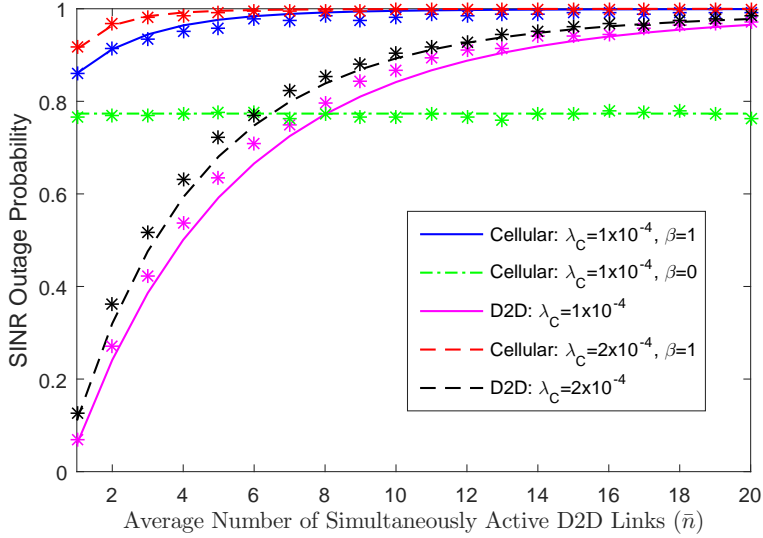}
  \caption{\small SINR outage probability as a function of average number of simultaneously active D2D links $\bar{n}$ for different values of cluster center density $\lambda_C$ ($\Gamma=40dB$). Simulation results are also plotted with markers. \normalsize}
\label{Fig_Navg_lambdaC}
\end{figure}

In Fig. \ref{Fig_Navg_lambdaC}, we plot the SINR outage probability of cellular and D2D links as a function of average number of simultaneously transmitting potential D2D UEs $\bar{n}$ in each cluster for different values of cluster center density $\lambda_C$. Moreover, the effect of spectrum sharing type is investigated. As described in Section \ref{sec:system_model}, $\beta$ indicates the type of spectrum sharing; i.e., it is equal to one for underlay and zero for the overlay scheme. For the underlay type of spectrum sharing, when the average number of simultaneously transmitting potential D2D UEs gets higher, both intra-cluster and inter-cluster interferences increase and as a result SINR outage probabilities for both cellular and D2D links increase. Similarly, inter-cluster interference increases with the increase in cluster center density. Therefore, outage probabilities for both cellular and D2D links increase. For the overlay type of spectrum sharing, outage probability is smaller for cellular UEs compared to underlay and it is independent of $\bar{n}$ since cross-mode interference from D2D UEs becomes zero in the case of overlay spectrum sharing. On the other hand, outage probability of D2D UEs remains the same with both overlay and underlay sharing, showing that the effect of cross-mode interference from cellular UEs is negligible even under the congested network scenario assumption.

\begin{figure}
\centering
  \includegraphics[width=\figsize\textwidth]{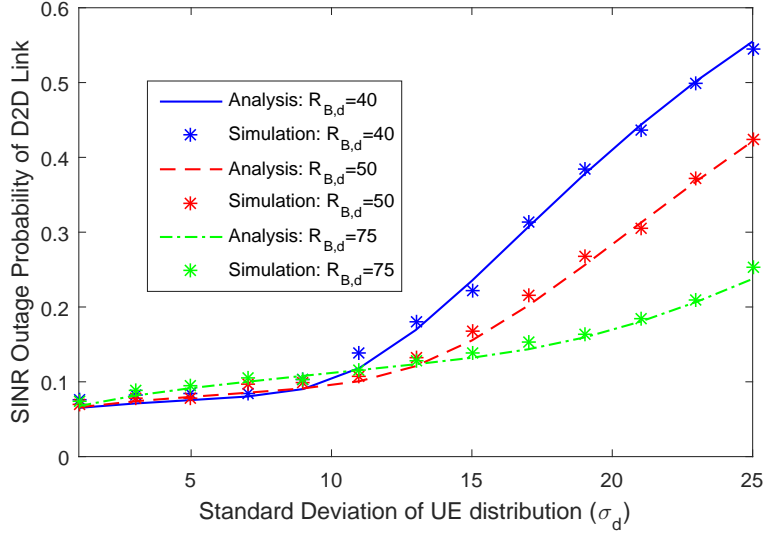}
  \caption{\small SINR outage probability as a function of UE distribution's standard deviation $\sigma_d$ for different  values of LOS ball radius $R_{B,d}$ ($\Gamma=20dB$). \normalsize}
\label{Fig_standard_deviation_RBd}
\end{figure}

In Fig. \ref{Fig_standard_deviation_RBd}, we investigate the effect of UE distribution's standard deviation $\sigma_d$ on SINR outage probability of D2D links for different values of LOS the ball radius $R_{B,d}$. We have two different observations depending on the value of $\sigma_d$. For small values of $\sigma_d$, i.e. when the potential D2D UEs in the cluster are distributed closer to each other, outage probability is less for small LOS ball radius $R_{B,d}$. On the other hand, outage probability with smaller LOS ball radius $R_{B,d}$ becomes greater for bigger values of $\sigma_d$. For small $\sigma_d$, main link is more likely be a LOS link and effect of interference is small if the LOS ball radius is small, hence outage probability is low. However, the main link becomes more likely to be a NLOS link and the effect of interference becomes relatively more dominant with the increasing $\sigma_d$.

\begin{figure}
\centering
  \includegraphics[width=\figsize\textwidth]{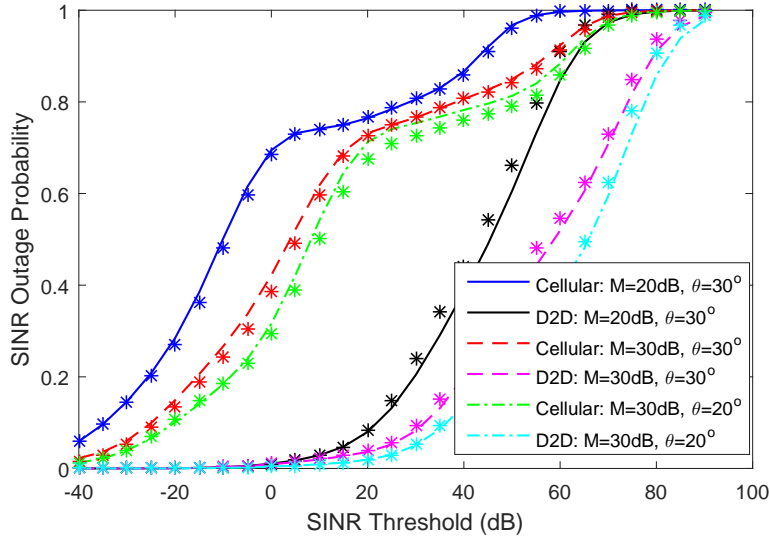}
  \caption{\small SINR outage probability as a function of the threshold in dB for different antenna parameters. Simulation results are also plotted with markers. \normalsize}
\label{Fig_antenna_pattern}
\end{figure}

Next, we compare the SINR outage probabilities for different values of the antenna main lobe gain $M_{\nu}$ and beam width of the main lobe $\theta_{\nu}$ for $\nu \in \{\BS, \UE\}$ in Fig. \ref{Fig_antenna_pattern}. Outage probability improves with the increase in the main lobe gain $M_{\nu}$ for the same value of $\theta_{\nu}$ for $\nu \in \{\BS, \UE\}$. On the other hand, since we assume perfect beam alignment for serving links, outage probability increases with the increase in the beam width of the main lobe due to the growing impact of the interference. Finally, we notice that for given SINR threshold, the outage probabilities for D2D links are smaller than those for cellular links, owing to generally smaller communication distances in D2D links.

\begin{figure}
\centering
  \includegraphics[width=\figsize\textwidth]{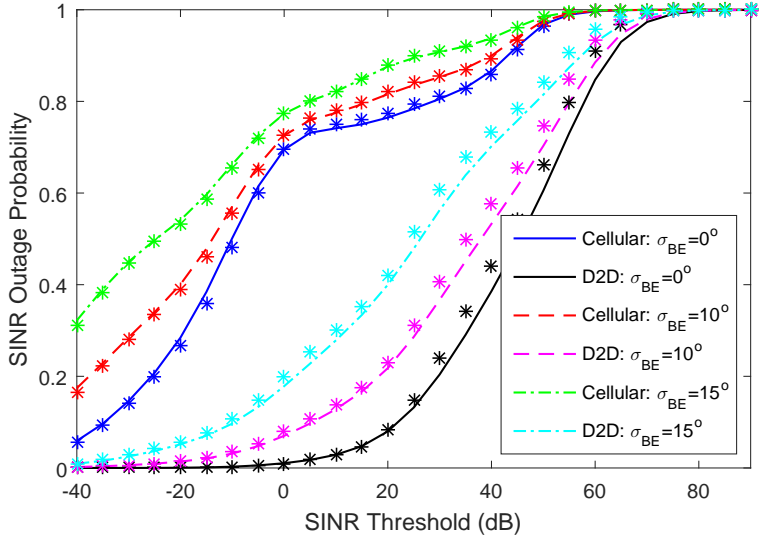}
  \caption{\small SINR outage probability as a function of the threshold in dB for different alignment errors
  $\sigma_{BE}$. Simulation results are also plotted with markers. \normalsize}
\label{Fig_BSE}
\end{figure}

Effect of beam steering errors between the transmitting nodes (cellular or potential D2D UEs) and receiving nodes (BSs or UEs) on the SINR outage probability of cellular and D2D links is shown  in Fig. \ref{Fig_BSE}. As shown in the figure, outage probability becomes worse for both cellular and D2D links with the increase in the standard deviation of the alignment error. Although the interference from interfering nodes remains unchanged, its effect grows with the increase in alignment error on the main link. This proves the importance of having perfect beam alignment to achieve improved performance.

\begin{figure}
\centering
  \includegraphics[width=\figsize\textwidth]{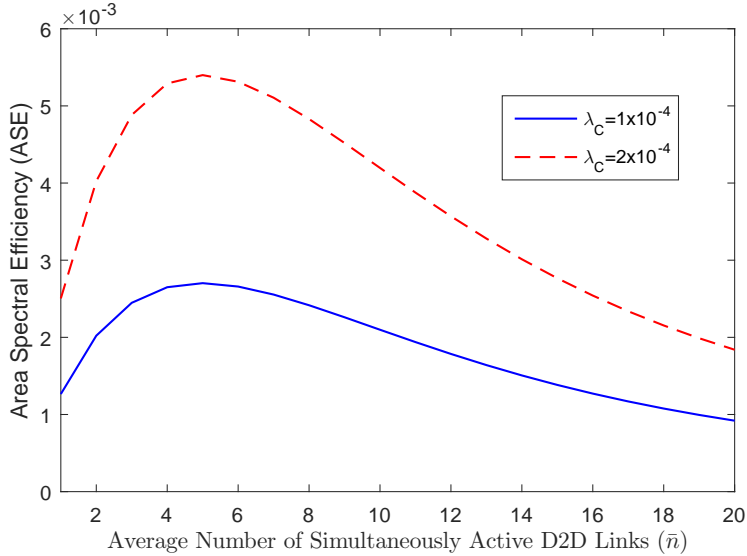}
  \caption{\small Area Spectral Efficiency (ASE) for underlay type of sharing as a function of average number of simultaneously active D2D links $\bar{n}$ for different values of cluster center density $\lambda_C$ ($\Gamma=40dB$). \normalsize}
\label{Fig_ASE_underlay}
\end{figure}

In the numerical analysis, we also investigate the ASE for underlay type of sharing. In Fig. \ref{Fig_ASE_underlay}, we plot the ASE as a function of the average number of simultaneously active D2D links $\bar{n}$ in each cluster for different values of the cluster center density $\lambda_C$. With the increase in the average number of simultaneously active D2D links, ASE first increases and then decreases. Therefore, an optimal value that maximizes ASE exists. Below this optimal value, increasing the average number of simultaneously active D2D links helps in improving spatial frequency reuse. Once the optimal value is exceeded, however, the effect of intra-cluster interference offsets the benefit of having larger average number of simultaneously active D2D links. Moreover, increasing cluster center density for the same average number of simultaneously active D2D links in each cluster improves ASE. Although the inter-cluster interference increases with larger cluster center density, spatial frequency reuse improves with larger cluster center density, i.e. $\bar{n}\lambda_C$ increases. Since inter-cluster interference does not have a dominant impact on outage probability and intra-cluster interference remains the same, ASE increases for the same average number of simultaneously active D2D links. Interestingly, the optimal number of simultaneously active D2D links is independent of the cluster center density because intra-cluster interference is more dominant than inter-cluster interference.

\begin{figure}
\centering
  \includegraphics[width=\figsize\textwidth]{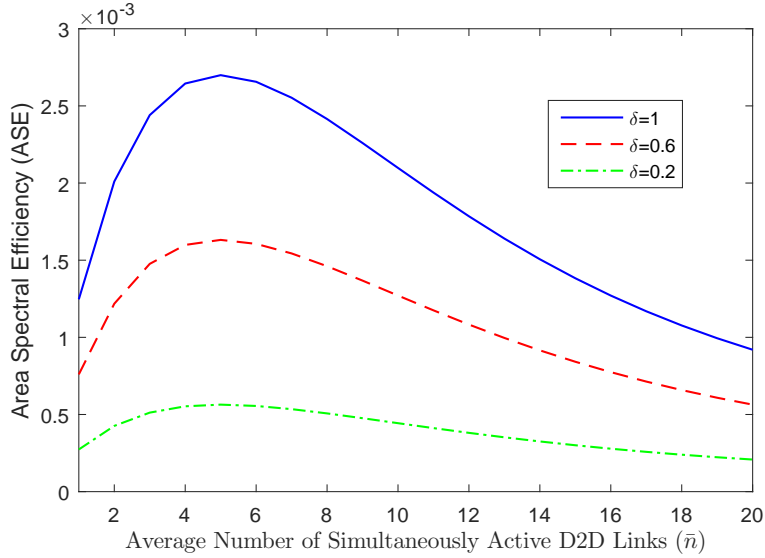}
  \caption{\small Area Spectral Efficiency (ASE) for overlay type of sharing as a function of average number of simultaneously active D2D links $\bar{n}$ for different values of spectrum partition factor $\delta$ ($\Gamma=40dB$). \normalsize}
\label{Fig_ASE_overlay}
\end{figure}

\begin{figure}
\centering
  \includegraphics[width=\figsize\textwidth]{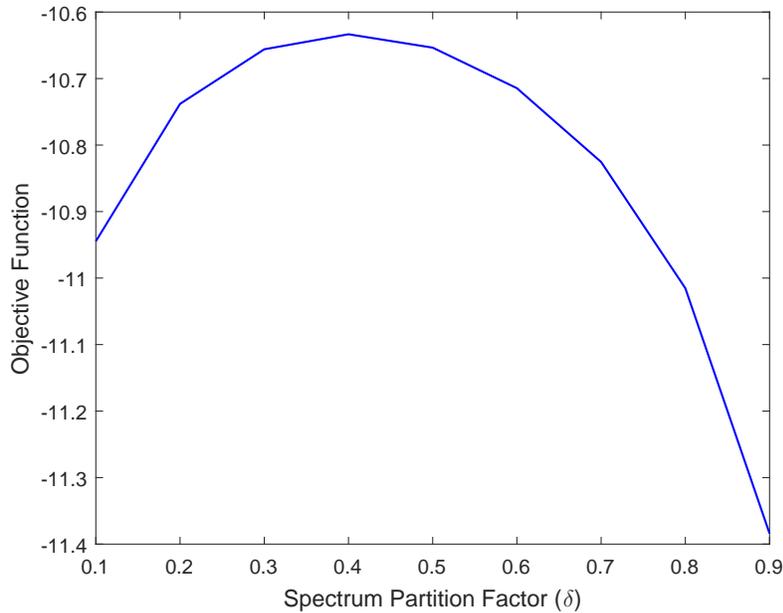}
  \caption{\small Objective function as a function of spectrum partition factor $\delta$ ($\Gamma=40dB$). \normalsize}
\label{Fig_obj_function}
\end{figure}

Finally, ASE is investigated for overlay type of spectrum sharing. In Fig. \ref{Fig_ASE_overlay}, ASE is plotted as a function of the average number of simultaneously active D2D links in each cluster $\bar{n}$ for different values of spectrum partition factor $\delta$. As  mentioned in Section \ref{sec:Analysis of Area Spectral Efficiency}, ASE is maximized if all bandwidth is assigned to D2D links, i.e. $\delta=1$. We also consider the optimal weighted proportional fair spectrum partition in Section \ref{sec:Analysis of Area Spectral Efficiency}, and plot the objective function in (\ref{eq:objection_func}) as a function of $\delta$. As shown in Fig. \ref{Fig_obj_function}, optimal spectrum partition factor is equal to $\delta^{\ast}=0.4=w_d$ which validates our result in Section \ref{sec:Analysis of Area Spectral Efficiency}.

\section{Conclusion} \label{sec:Conclusion}
In this paper, we have provided an analytical framework to compute the SINR outage probabilities for both cellular and D2D links in a D2D-enabled mmWave cellular network with clustered UEs. Distinguishing features of mmwave communications, such as directional beamforming with sectored antenna model and modified LOS ball model for blockage modeling, have been considered in the analysis. BSs and cellular UEs are assumed to be distributed according to independent PPPs, while potential D2D UEs locations' are modeled as a PCP. Potential D2D UEs in the clusters are allowed to choose cellular or D2D mode according to a flexible mode selection scheme. Under these assumptions, we have analyzed the interference experienced in cellular uplink and D2D links, and characterized the SINR outage probabilities.

Numerical results show that probability of selecting D2D mode decreases with increasing UE distribution's standard deviation $\sigma_d$ and increasing $p_{L,c}$, while increase in $p_{L,d}$ leads to higher D2D mode selection probability. We have also shown that more simultaneously transmitting potential D2D UEs and/or higher cluster center density result in higher outage probabilities for both cellular and D2D links due to the growing impact of interference. Moreover, the type of spectrum sharing plays a crucial role in the SINR outage performance of cellular UEs. Another interesting observation is that smaller LOS ball radius is preferred for small values of $\sigma_d$ while the opposite is advantageous for large values of $\sigma_d$. Moreover, increasing the main lobe gain and decreasing the beam width of the main lobe result in lower SINR outage. Effect of alignment error on outage probability is also quantified and importance of beam alignment in improving the performance is noted. Finally, ASE of the whole network is analyzed for both underlay and overlay types of sharing. We have shown that there is an optimal number of simultaneously active D2D links maximizing ASE. This optimal number is independent of cluster center density and spectrum partition factor. For overlay, there exists an optimal spectrum partition factor if the optimal weighted proportional fair spectrum partition is considered. Analyzing the outage performance of cellular and D2D UEs for a different PCP such as uniformly distributed potential D2D UEs, and investigating the effect of using different mode selection schemes remains as future work.

\appendix
\subsection{Proof of Lemma 1} Probability of selecting the D2D mode for a potential D2D UE located in a cluster $x \in \Phi_C$ can be computed as
\label{Proof of Lemma 1}
\begin{align}
P_{D2D}&=\sum_{s \in \{L,N\}} \sum_{s^{\prime} \in \{L,N\}} \mathbb{P}\left (T_d r_d^{-\alpha_{s^{\prime},d}} \geq  r_c^{-\alpha_{s,c}}\right) p_{s^{\prime},d}(r_d) \mathcal{B}_{s,c}  \nonumber \\
&= \sum_{s \in \{L,N\}} \sum_{s^{\prime} \in \{L,N\}}\mathbb{P}\left (r_c \geq r_d^{\alpha_{s^{\prime} ,d} /\alpha_{s,c}}T_d^{-1/\alpha_{s,c}}\right ) p_{s^{\prime},d}(r_d) \mathcal{B}_{s,c} \nonumber \\
&=\sum_{s \in \{L,N\}}  \sum_{s^{\prime} \in \{L,N\}} \int_0^{\infty} \int_0^{\infty} \bar{F}_s\left(\frac{r_d^{\alpha_{s^{\prime},d}/\alpha_{s,c}}}{T_d^{1/\alpha_{s,c}}}\right) f_{R_d}(r_d|\omega) f_{\Omega}(\omega) p_{s^{\prime},d}(r_d)\mathcal{B}_{s,c} dr_d d\omega \nonumber \\
&\stackrel{(a)}{=}\sum_{s \in \{L,N\}}  \sum_{s^{\prime} \in \{L,N\}} \int_0^{\infty} \int_0^{\infty}  e^{-2\pi\lambda_B \psi_s\left(r_d^{\alpha_{s^{\prime},d}/\alpha_{s,c}}/T_d^{1/\alpha_{s,c}}\right)} f_{R_d}(r_d|\omega) f_{\Omega}(\omega) p_{s^{\prime},d}(r_d) dr_d d\omega
\end{align}
where $\bar{F}_s(r_c)=e^{-2\pi\lambda_B\psi_s(r_c)}/\mathcal{B}_{s,c}$ is the complementary cumulative distribution function (ccdf) of the cellular link distance $r_c$ to the nearest LOS/NLOS BS, $p_{s^{\prime},d}(r_d)$ is the LOS/NLOS probability function for the D2D link given in (\ref{LOS_prob_funct}), and (a) follows by substituting the cdf of $r_c$ into the expression.

\subsection{Proof of Lemma 2}
\label{Proof of Lemma 2}
Laplace transform of the aggregate interference at the BS from cellular UEs transmitting in the same uplink channel in different cells can be calculated using (\ref{eq:LT}) as follows:
\begin{align}
\mathcal{L}_{I_{cc}}(v)= \prod_{G} \prod_j \mathcal{L}_{I_{cc,j}^G}(v), \label{eq:LT_cc6}
\end{align}
where the Laplace transform for $I_{cc,j}^G$ can be computed using stochastic geometry as follows:
\begin{align}
\mathcal{L}_{I_{cc,j}^G}(v)&\stackrel{(a)}{=} \exp\left(-2\pi\lambda_{B}p_{G} \int_{0}^{\infty}\left(1-\mathbb{E}_h \left[ e^{-v P_{c}G h t^{-\alpha_{j,c}} }\right]\right)Q(t^{\alpha_{j,c}}) p_{j,c}(t) t dt\right) \nonumber \\
&\stackrel{(b)}{=} \exp\left(-2\pi\lambda_B p_{G} \int_{0}^{\infty} \frac{v P_cG t^{-\alpha_{j,c}}}{1+v P_cG t^{-\alpha_{j,c}}}Q(t^{\alpha_{j,c}})p_{j,c}(t)tdt\right), \label{eq:LT_cc_j}
\end{align}
where (a) follows from computing the probability generating functional (PGFL) of PPP, the lower limit of integration is determined using the  fact that the minimum distance of interfering UEs in cellular mode to the typical BS is equal to $r_c$, (b) is obtained by computing the moment generating function (MGF) of the exponentially distributed random variable $h$, and $Q(y)$ is given in (\ref{Q(y)}). By inserting (\ref{eq:LT_cc_j}) into (\ref{eq:LT_cc6}) for $j \in \{L,N\}$ and $G \in \{M_{\BS}M_{\UE},M_{\BS}m_{\UE}, m_{\BS}M_{\UE},m_{\BS}m_{\UE}\}$, the Laplace transform expression in Lemma 2 can be obtained.

\subsection{Proof of Lemma 3}
\label{Proof of Lemma 3}
Laplace transform of the aggregate interference at the BS from both intracell and intercell D2D UEs can be calculated using (\ref{eq:LT})
\begin{align}
\mathcal{L}_{I_{dc}}(v)= \prod_G \prod_j \mathcal{L}_{I_{dc,j}^G}(v), \label{eq:LT_dc6}
\end{align}
where the Laplace transform for $I_{dc,j}^G$ can be computed using stochastic geometry and following the similar steps as in \cite{Afshang}:
\begin{align}\label{eq:LT_dc_j}
&\mathcal{L}_{I_{dc,j}^G}(v) \nonumber \\
 &\stackrel{(a)}{=}\mathbb{E}_{\Phi_C}\left[\prod_{x \in \Phi_C} \mathbb{E}_{\mathcal{A}_d^{x}} \left[\prod_{y \in \mathcal{A}_d^{x}} \mathbb{E}_{h_{y_x}}\left[ e^{-vP_dGh_{y_x}r_{y_x}^{-\alpha_{j,d}}} \right] \right] \right] \nonumber \\
  &\stackrel{(b)}{=}\mathbb{E}_{\Phi_C}\left[\prod_{x \in \Phi_C} \mathbb{E}_{\mathcal{A}_d^{x}} \left[\prod_{y \in \mathcal{A}_d^{x}} \frac{1}{1+vP_dGr_{y_x}^{-\alpha_{j,d}}}\right] \right]  \nonumber \\
  &\stackrel{(c)}{=}\mathbb{E}_{\Phi_C}\left[\prod_{x \in \Phi_C} \sum_{k=0}^{N/2}\left(\int_{\mathbb{R}^2}  \frac{1}{1+vP_dGr_{y_x}^{-\alpha_{j,d}}} p_{j,d}(r_{y_x})f_{Y}(y)dy \right)^k \mathbb{P}(K=k|K<N/2)\right]  \nonumber \\
  &\stackrel{(d)}{=}\exp\left( -\lambda_C \int_{\mathbb{R}^2}\left(1-\sum_{k=0}^{N/2}\left(\int_{\mathbb{R}^2}  \frac{1}{1+vP_dGr_{y_x}^{-\alpha_{j,d}}} p_{j,d}(r_{y_x})f_{Y}(y)dy \right)^k \frac{(\bar{n}P_{D2D})^ke^{-(\bar{n}P_{D2D})}}{k!\eta} \right)dx \right) \nonumber \\
  &\stackrel{(e)}{=}  \exp \left( -2\pi\lambda_C \int_0^{\infty}\left(1 -\sum_{k=0}^{N/2} \left(\int_{0}^{\infty} \frac{1}{1+vP_dGu^{-\alpha_{j,d}}}p_{j,d}(u)f_U(u|t)du\right)^k \frac{(\bar{n}P_{D2D})^ke^{-(\bar{n}P_{D2D})}}{k!\eta}  \right) tdt \right)     \nonumber \\
  &\stackrel{(f)}{=} \exp \left( -2\pi\lambda_C \int_0^{\infty}\left(1 -\exp\left(-\bar{n}P_{D2D} \int_{0}^{\infty} \frac{vP_dGu^{-\alpha_{j,d}}}{1+vP_dGu^{-\alpha_{j,d}}}p_{j,d}(u)f_U(u|t)du \right) \right) tdt \right)
\end{align}
where $\{r_{y_x}=\|x+y\|, \forall x \in \Phi_C, \forall y \in \mathcal{A}_d^{x}\}$, (a) follows from the assumption of independent fading gains across all interfering links, (b) is obtained by computing the moment generating function (MGF) of the exponentially distributed random variable $h_{y_x}$, (c) follows from the fact that the locations of the cluster members in each cluster are independent when conditioned on $x \in \Phi_C$ and expectation over the number of interfering devices which are Poisson distributed conditioned on the total being less than $N/2$ where $\eta=\sum_{l=0}^{N/2}\frac{(\bar{n}P_{D2D})^le^{-(\bar{n}P_{D2D})}}{l!}$, (d) is determined by computing the probability generating functional (PGFL) of PPP, (e) follows by converting the coordinates from Cartesian to polar, (f) follows from the assumption that $\bar{n}P_{D2D} \ll N/2$.  By inserting (\ref{eq:LT_dc_j}) into (\ref{eq:LT_dc6}) for $j \in \{L,N\}$ and $G \in \{M_{\BS}M_{\UE},M_{\BS}m_{\UE}, m_{\BS}M_{\UE},m_{\BS}m_{\UE}\}$, we obtain the Laplace transform expression in Lemma 3.

\subsection{Proof of Lemma 5}
\label{Proof of Lemma 5}
Laplace transform of the intra-cluster interference at the typical UE $\in \mathcal{N}_r^{x_0}$ in the representative cluster can be calculated using (\ref{eq:LT}) as follows:
\begin{align}
\mathcal{L}_{I_{dd_{\text{intra}}}}(v|w_0))= \prod_G \prod_j \mathcal{L}_{I_{dd_{\text{intra}},j}^G}(v), \label{eq:LT_ddintra6}
\end{align}
where the Laplace transform for $I_{dd_{\text{intra}},j}^G$ conditioned on $w_0$ can be computed following similar steps as in the proof of Lemma 3:
\begin{align}\label{eq:LT_ddintra_j}
 \mathcal{L}_{I_{dd_{\text{intra}},j}^G}(v) &= \mathbb{E}_{\mathcal{A}_d^{x_0}} \left[\prod_{y \in \mathcal{A}_d^{x_0} \setminus y_0} \mathbb{E}_{h_{y_{x_0}}}\left[ e^{-vP_dGh_{y_{x_0}}r_{d1}^{-\alpha_{j,d}}} \right] \right] \nonumber \\
&\stackrel{(a)}{=} \mathbb{E}_{\mathcal{A}_d^{x_0}} \left[\prod_{y \in \mathcal{A}_d^{x_0} \setminus y_0} \frac{1}{1+vP_dGr_{d1}^{-\alpha_{j,d}}}\right] \nonumber \\
&\stackrel{(b)}{=} \sum_{k=0}^{N/2-1}\left(\int_{\mathbb{R}^2}  \frac{1}{1+vP_dGr_{d1}^{-\alpha_{j,d}}} p_{j,d}(r_{d1})f_{Y}(y)dy \right)^k \mathbb{P}(K=k|K<N/2-1)  \nonumber \\
&=\sum_{k=0}^{N/2-1}\left(\int_{\mathbb{R}^2}  \frac{1}{1+vP_dGr_{d1}^{-\alpha_{j,d}}} p_{j,d}(r_{d1})f_{Y}(y)dy \right)^k \frac{(\bar{n}P_{D2D}-1)^ke^{-(\bar{n}P_{D2D}-1)}}{k!\eta} \nonumber \\
&\stackrel{(c)}{=} \exp \left(-\left(\bar{n}P_{D2D}-1\right)  \int_0^{\infty} \frac{vP_d G u^{-\alpha_{j,d}}}{1+vP_dG u^{-\alpha_{j,d}}}f_{U}(u|w_0)p_{j,d}(u)du \right)
\end{align}
where $\{r_{d1}=\|x_0+y\|, \forall y \in \mathcal{A}_d^{x_0} \setminus y_0\}$, $\eta=\sum_{l=0}^{N/2-1}\frac{(\bar{n}P_{D2D}-1)^le^{-(\bar{n}P_{D2D}-1)}}{l!}$, (a) is obtained by computing the moment generating function (MGF) of the exponentially distributed random variable $h_{y_{x_0}}$, (b) follows from the fact that the locations of the intra-cluster D2D UEs simultaneously transmitting in D2D mode are independent when conditioned on $x_0 \in \Phi_C$ and expectation over the number of interfering devices which are Poisson distributed conditioned on the total being less than $N/2-1$, (c) follows by converting the coordinates from Cartesian to polar and using the assumption that $\bar{n}P_{D2D} \ll N/2$. By inserting (\ref{eq:LT_ddintra_j}) into (\ref{eq:LT_ddintra6}) for $j \in \{L,N\}$ and $G \in \{M_{\UE}M_{\UE},M_{\UE}m_{\UE}, m_{\UE}M_{\UE},m_{\UE}m_{\UE}\}$, we readily obtain the Laplace transform expression in Lemma 4.

\subsection{Proof of Theorem 1}
The outage probability for a typical UE in cellular mode can be calculated as follows:
\label{Proof of Theorem 1}
\begin{align}
&\Pout^c(\Gamma)= \PoutL^c(\Gamma)\mathcal{B}_{L,c}+\PoutN^c(\Gamma)\mathcal{B}_{N,c} \nonumber \\
&\Pout^c(\Gamma)=\sum_{s \in \{L,N\}}\mathbb{P}\left( \frac{P_c G_0h_0r_c^{-\alpha_{s,c}}}{\sigma^2+I_{cc}+I_{dc}} \leq \Gamma \right) \mathcal{B}_{s,c} \nonumber \\
&=\sum_{s \in \{L,N\}} \int_0^{\infty}\hspace{-0.2cm} \mathbb{P}\left (h_0 \leq \frac{\Gamma r_c^{\alpha_{s,c}}}{P_cG_0}\left(\sigma^2+I_{cc}+I_{dc}\right) |r_c \right)f_{s}(r_c) \mathcal{B}_{s,c} dr_c \nonumber \\
&= 1- \sum_{s \in \{L,N\}} \int_0^{\infty} e^{-v\sigma^2} \mathcal{L}_{I_{cc}}(v) \mathcal{L}_{I_{dc}}(\beta v) f_{s}(r_c) \mathcal{B}_{s,c} dr_c \label{OutageProbability}
\end{align}
where $v=\frac{\Gamma r_c^{\alpha_{s,c}}}{P_{c}G_0}$, and (\ref{OutageProbability}) follows from $h_0$ $\sim$ $\exp(1)$, and by noting that Laplace transforms of interference at the BS from cellular UEs and D2D UEs are independent.

\end{spacing}
\vspace{-.2cm}

\begin{spacing}{1.5}

\end{spacing}
\end{document}